\documentclass[]{spie}  

\usepackage[numbers]{natbib}
\usepackage{aas_macros}
\usepackage{amsmath,amsfonts,amssymb}
\usepackage{graphicx}
\usepackage[colorlinks=true, allcolors=blue]{hyperref}

\title{Sensitivity and Performance of LBTI/NOMIC Spectroscopy: Prospects for 10-- and 30-- meter class Mid-IR Exoplanet Science}

\author[a,b]{Brittany E.~Miles} 
\author[a,c]{Steve Ertel} 
\author[a]{Kevin Wagner} 
\author[a,d]{Daniel Apai}

\affil[a]{Department of Astronomy and Steward Observatory, The University of Arizona, 933 N Cherry Ave., Tucson, AZ 85721, USA}
\affil[b]{51 Pegasi b Fellow}
\affil[c]{Large Binocular Telescope Observatory, The University of Arizona, 933 N Cherry Ave., Tucson, AZ 85721, USA}
\affil[d]{Lunar and Planetary Laboratory, The University of Arizona, 1640 E. University Blvd., Tucson, AZ 85721, USA}

\authorinfo{Further author information: (Send correspondence to B.E.M.)\\B.E.M.: E-mail: bemiles@arizona.edu}

\begin{document} 
\maketitle

\begin{abstract}
Long wavelength infrared (8-13 $\mu$m) spectroscopy is invaluable for detecting molecular features in the atmospheres of gas giant and terrestrial exoplanets. The nulling-optimized mid-infrared camera (NOMIC) on the Large Binocular Telescope Interferometer (LBTI) has a low resolution (R$\sim$200) germanium grism that was previously installed but has not been characterized and commissioned for scientific observations. Using a 1.27 mm slit and broadband filter in combination with the grism, the infrared window between 8-13 $\mu$m can be captured. We describe initial on sky testing of the LBTI/NOMIC grism mode with adaptive optics to study standard stars and binaries. We discuss the impact of observational strategy and telluric calibration on the spectral reduction process. We infer the impact of existing mid-infrared detectors on NOMIC's spectroscopic mode and discuss requirements to enable higher resolution 8-13 $\mu$m spectroscopy on current and future facilities.
\end{abstract}


\section{INTRODUCTION}
\label{sec:intro}

\begin{figure} [ht]
   \begin{center}
   \begin{tabular}{c} 
   \includegraphics[width=6.5in]{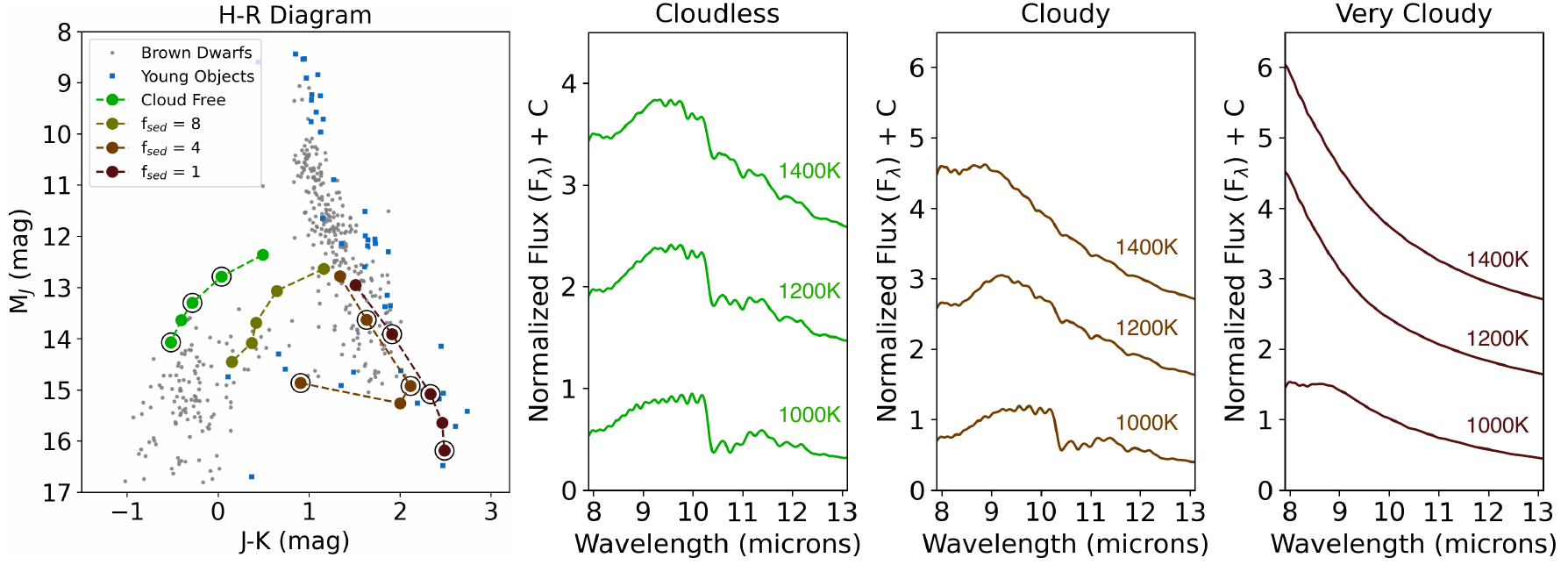}
   \end{tabular}
   \end{center}
   \caption[example]{ \label{fig:HR_diagram_and_spectra}Clouds have a massive impact on long-wavelength spectra and near-infrared color. The left plot shows an H-R Diagram of brown dwarfs and directly imaged exoplanets from the UltracoolSheet(\citenum{best_2024_10573247}). Brown dwarfs are plotted as grey dots. Young objects that may include directly imaged exoplanets are plotted as blue squares. The photometry of a few atmospheric models (Sonora Diamondback, \citenum{2024arXiv240200758M}) with a fixed surface gravity of log(g) = 5, are plotted as large green/brown circles. Models with effective temperatures of 1600K, 1400K, 1200K, 1100K, and 1000K are plotted with four different cloudiness parameters: No clouds, f$_{\mathrm{sed}}$ = 8, 4, and 1. Circled model photometry have their spectra from 8-13 $\mu$m plotted in the right three panels at R $\sim$ 200.}
   \end{figure}

Within the next decade, extremely large telescopes will come online creating a new opportunity for mid-infrared instruments that operate at small inner working angles ($<$.2 $^{\prime\prime}$) to study planet atmospheres, planetary formation, and planetary system evolution. The 8-13 $\mu$m region is vital for detecting and characterizing planets of many classes. At low spectral resolutions (R$\sim$20), biosignatures such as ozone (O$_{3}$) in Earth-like planets are detectable (\citenum{2007ApJ...658..598K, 2020ApJ...898L..42M}). Nitrogen bearing molecules and forests of hydrocarbons are easily distinguished within the atmospheres of cooler gas giant exoplanets and Solar System gas giants (\citenum{2006ApJ...648..614C, 2022SPIE12184E..1UB, 2023ApJ...951L..48B, 2023Natur.624..263B, 2023RemS...15.1811R}). 
The presence, onset temperature, and grain properties of silicate clouds in warmer gas giant exoplanets can also be directly explored with low and medium resolution spectroscopy (\citenum{2006ApJ...648..614C, 2021ApJ...920..146L, 2022MNRAS.513.5701S, 2023MNRAS.523.4739S, 2023ApJ...946L...6M}, Figure~\ref{fig:HR_diagram_and_spectra}). Mid-infrared spectroscopy is informative, but also time expensive and sensitive to calibration errors primarily due to Earth's bright sky background at these wavelengths. Successful and efficient mid-infrared spectroscopy at small inner working angles on 30-meter class telescopes requires investment in developing observational strategies and new technologies on current facilities. 

The Large Binocular Telescope Interferometer (LBTI, \citenum{2016SPIE.9907E..04H, 2020SPIE11446E..07E}) contains the nulling-optimized mid-infrared camera (NOMIC, \citenum{2014SPIE.9147E..1OH}), which is used as an 8-13 $\mu$m imager. NOMIC resides in the Nulling and Imaging Cryostat (NIC) with two other instruments, LBT Mid-infrared Camera (LMIRCam, \citenum{2010SPIE.7735E..3HS}) and PhaseCam (\citenum{2014SPIE.9146E..09D, 2020JATIS...6c5001M}). Light collected from the primary mirrors of the Large Binocular Telescope (LBT) is sent through the universal beam combiner then fed into NIC. Within NIC, a trichroic diverts the 1-2.5 $\mu$m and 8-13 $\mu$m light into NOMIC. To use NOMIC's standard imaging mode, another imaging dichroic is used to isolate the 8-13 $\mu$m light into the focal plane on the detector. Just before light reaches the NOMIC detector, it passes a pupil wheel, followed by two filter wheels. To enable NOMIC spectroscopy with the standard imaging optical path, a slit is placed into the pupil plane followed by filter in wheel 1 and a grism in filter wheel 2\footnote{LBTI and NOMIC's optical layout -- Figures 3.8 and 3.16 in https://scienceops.lbto.org/lbti/lbti-users-manual/}.

A germanium grism was initially installed in NOMIC to get dispersed interference fringe measurements, but it was not implemented due to complexity and low signal to noise (Philip Hinz, private communication). NOMIC's grism is a fabrication residual from grisms developed for LMIRCam for slitless spectroscopy (\citenum{2012SPIE.8450E..3PK}). The grism's substrate germanium transmits between 8-13 $\mu$m but was not optimized for transmission over this wavelength range (\citenum{2012SPIE.8450E..3PK}). Resolution estimates of these grisms were only completed at L-- and M-- band.  Over the Spring 2023 and Fall 2023 observing semesters we completed dome and on-sky testing of the grism to characterize its spectral grasp, wavelength resolution, sensitivity, and spatial resolution. A set of standard stars were observed to validate NOMIC's ability to measure known spectral responses. Two binary systems were observed to understand how well NOMIC could spatially resolve two dispersed point sources. 

We summarize the observational strategies and results of these tests in this paper. In Section~\ref{sec:observations}, we describe the observations taken and then discuss the data reduction in Section~\ref{sec:data reduction}. In Section~\ref{sec:mode performance}, the initial performance and sensitivity of NOMIC derived from on-sky observations are outlined. We also describe the impact of airmass on telluric calibration. Section~\ref{sec:discussion} discusses future plans for science observations, the impact of planned upgrades to NOMIC, and detector requirements to enable higher resolution modes at the Large Binocular Telescope and the Giant Magellan Telescope.

\section{Observations}\label{sec:observations}

\subsection{Dome Testing}
\begin{figure} [ht]
   \begin{center}
   \begin{tabular}{c} 
   \includegraphics[width=6.5in]{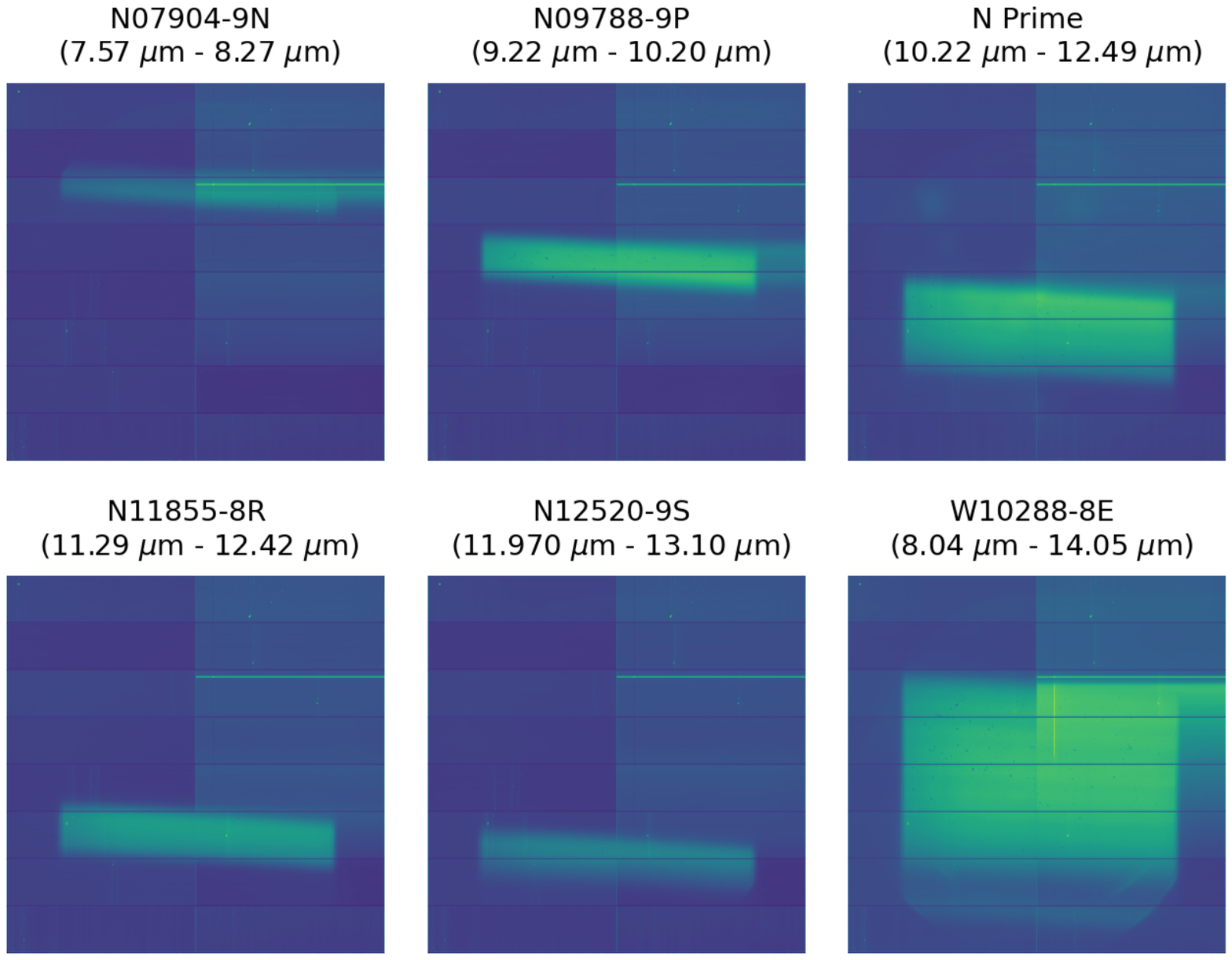}
   \end{tabular}
   \end{center}
   \caption[example]{\label{fig:footprint}Raw detector images of telescope dome observations executed in NOMIC's spectroscopic mode. We tested six different filters in combination with the grism. The filter names and their wavelength coverage are listed above each detector image. The W10288-8E filter has the broadest coverage and is used for science observations.}
\end{figure}

On March 11, 2023, the spectral footprint of LBTI/NOMIC was estimated by pointing the telescope at the dome. NOMIC was set up with the 1.27 mm slit in the pupil wheel followed by a filter in filter wheel 1 and the germanium grism in filter wheel 2. The NOMIC detector was used in high gain mode with full frame (1024 x 1024) readout.  An integration time of 0.247 seconds was used to avoid saturation on the detector. Several filters with known responses were tested in combination with the grism to assess the spectral grasp of the instrument mode shown in Figure~\ref{fig:footprint}. The W10288-8E filter provides the broadest grasp covering 8-14 $\mu$m and is used for science observations. The remaining filters were used in future observations to try validating the wavelength solution.

There are several cosmetics and detector effects that impact the easily usable science area of the mode. The NOMIC detector is a Raytheon Aquarius detector split into two columns of eight channels. The boundaries between the columns and channels are not light sensitive creating 1 to 2 pixel wide regions of little to no response on the detector. Based on dome observations, the central column can be avoided by nodding a target in the x-direction on the detector. The channel boundaries are unavoidable in the spectral direction. The right column of the NOMIC detector has noisier channels, therefore the left column is preferred for nodding point sources during science observations. Cold, circular vignetting with a radius of $\sim$15 arcsec impacts the left-most and right-most $\sim$1.5 arcsec of the detector area. The slit has a length of $\sim$13.5 arcsec centered near horizontal in the field of view.

\subsection{On Sky Testing}
\begin{figure}[h]
   \begin{center}
   \begin{tabular}{c} 
   \includegraphics[width=6.5in]{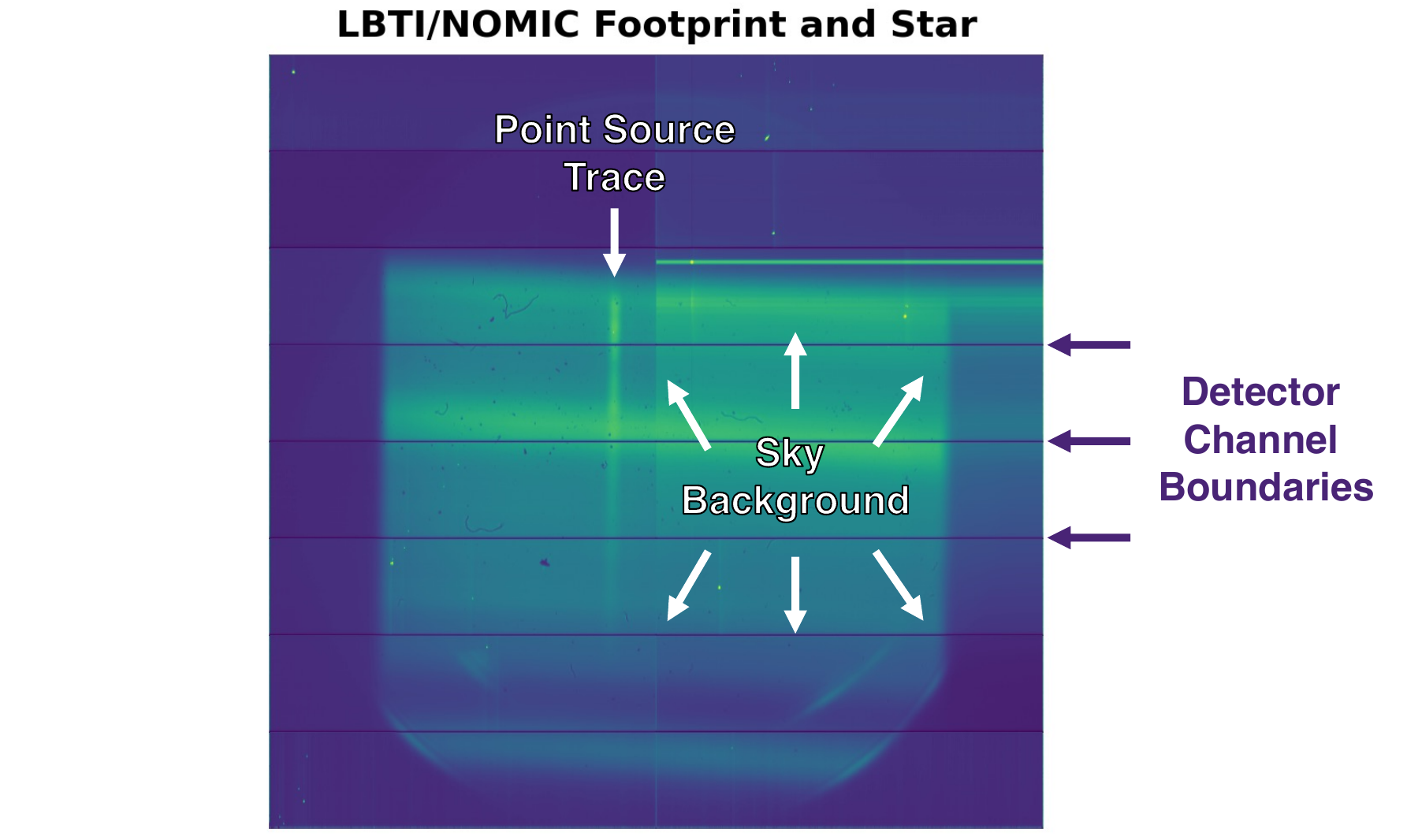}
   \end{tabular}
   \end{center}
   \caption[example]{\label{fig:science_image}Example of an on-sky science exposure from NOMIC's spectroscopic mode, which uses the W10288-8E filter in series with the grism. A point source is placed into the slit and Earth's sky background fills the entire slit area. When dispersed, the point source creates a trace. The sky background is also dispersed but over the full extent of the slit in the spatial direction. The 1 - 2 pixel wide horizontal detector channel boundaries are highlighted with blue arrows. \\}
\end{figure}

We tested NOMIC's spectroscopic mode during one night in September 2023 and three nights in December 2023. The 2023-Sep-24 observing run was used to observe very bright A-type stars to validate the mode's ability to measure a blackbody response. An example science image with a point source is shown in Figure~\ref{fig:science_image}. The September observations were completed in relatively poor conditions with high water vapor column density and poor seeing. The December run measured the effect of varying air mass and attempted to measure the spatial resolution of the mode on binary systems. The 2023-Dec-26 and 2023-Dec-27 observations had worse conditions with pockets of high and unstable seeing. The 2023-Dec-29 observations had high water vapor, but stable seeing. Precipitable water vapor (PWV) is estimated at the Submillimeter Telescope Observatory (SMT) on Mount Graham using the a 225 GHz radiometer. No radiometer data was taken during any of the December dates, but the sky background seen in other instruments that night was bright enough to pause regular mid-infrared science observations.

On 2023-Sep-24 we took spectra of Lambda Persei (V mag = 4.29, K mag = 4.14) and Sirius (V mag = -1.46, K mag =  -1.35) \footnote{V mags reported from SIMBAD, \url{http://simbad.cds.unistra.fr/simbad/}}. The targets were located using LBTI/LMIRCam as an acquisition camera and then centered in NOMIC's slit. Spectroscopic observations were taken by nodding the target between two positions along the slit. The two nod positions are 2$^{\prime\prime}$ apart and placed on the left half of the detector where there is less detector noise. One second integrations were used to minimize detector saturation. The first two nod pairs of Lambda Persei used 300 exposures per nod, totaling to about five minutes per nod pair. In the last three nod pairs, the exposures per nod were reduced to 60 or one minute per nod pair for better background subtraction. Two sets of 1-minute nods pairs were taken for Sirius. Seeing was relatively stable at 1$^{\prime\prime}$ over the observations, however precipitable water vapor changed by 2mm over the night. These observations are summarized in Table~\ref{tab:September_Run}.

The 2023-Dec-26 and 2023-Dec-27 observations were used to collect spectra and images of the binary systems HD 81212 (primary: V mag 6.7, K mag =  5.53) and HD 10453 (primary: V mag = 5.8, K mag =  4.60, primary) respectively. NOMIC's slit does not rotate and always stays near horizontal to the horizon. Binaries need to be observed when the system's combined positional angle and parallactic angle is near 90$^{\circ}$ or 270$^{\circ}$ to maximize the separation of  the spectral traces on the NOMIC detector. The HD 81212 AB system has a position angle of 94$^{\circ}$ with a separation of .75$^{\prime\prime}$ (\citenum{1999AJ....118.1395D}). HD 10453 AB has a position angle of 316.1$^{\circ}$ with a separation of 1.87$^{\prime\prime}$ (\citenum{1999AJ....118.1395D}). NOMIC was set up to take 7 nod pairs of HD 81212. Each nod is composed of 60 exposures that are 0.98825 seconds long. Alpha Leo was used as the calibrator star and 2 nod pairs, with the same settings were taken. The same set up was used for HD 10453 to take 8 nod pairs with Alpha Piscium as the calibrator target. Only the brighter star or a combination of the two was easily visible in initial reductions of both systems. These observations suffered from poor seeing that varied on timescales shorter than the nods times. A summary of the binary observations are documented in Tables~\ref{tab:December_Run_2} and \ref{tab:December_Run_3}.

A-- and B-- spectral type star observations were done again on 2023-Dec-29. The three targets were Alpha Coronae Borealis (A1, V mag = 2.24, K mag = 2.20), Beta Ursae Majoris (A1, V mag = 2.37, K mag = 2.35), and Alpha Leo (B8, V mag = 1.40 , K mag =  1.62). The same 1-minute nod set up was repeated for these targets. Alpha Coronae Borealis had 7 nod pairs taken, Beta Ursae Majoris had 7 nod pairs taken, and Alpha Leo had 6 nod pairs taken. More nod pairs were taken than usual to build signal at longer wavelengths. These observations were taken in poor water vapor conditions, but had stable seeing for each target. These observations are summarized in Table~\ref{tab:December_Run_4}.

\begin{table}[ht]
\caption{Summary of A stars observed for the first on sky run on 2023-Sep-24. Nod pairs are used to subtract background from each other. The second block of Lambda Persei observations are not used in the reduction because of inconsistent background between both nods.} 
\label{tab:September_Run}

\begin{center}       
\begin{tabular}{l|c|c|c|c|c|c|c|c} 
\textbf{Target} & SpT & Nods       & Exposures & Integration Time & Total Exposure & PWV & AM & seeing \\
\textbf{Name}   &      &            & per Nod   & per Exposure (s) & Time (s)       & (mm)    &     & (arcsec)         \\ \hline
Lambda Persei   &  A0      & 2         & 300       & 0.98825    & 592.95 & 5.58  & 1.05 & 0.97  \\
Lambda Persei*   &  A0      & 2         & 300       & 0.98825          & 592.95 & 4.29  & 1.05 & 1.00  \\
Lambda Persei   &  A0      & 2         & 60        & 0.98825          & 118.59  & 4.84  & 1.05 & 1.01  \\ 
Lambda Persei   &  A0      & 2         & 60        & 0.98825          & 118.59  & 4.54  & 1.05 & 1.01  \\
Lambda Persei   &  A0      & 2         & 60        & 0.98825          & 118.59  & 5.06  & 1.05 & 1.04  \\
\hline
Sirius          &  A1      & 2         & 60        & 0.98825          & 118.59 & 6.21  & 1.84 & 1.05  \\
Sirius          &  A1      & 2         & 60        & 0.98825          & 118.59 & 5.63  & 1.82 & 1.01  \\

\end{tabular}
\end{center}
\end{table}

\begin{table}[ht]
\caption{Summary of targets observed for the first on sky run on 2023-Dec-26. HD 81212 is a binary, but the spectral type of the primary is listed. Precipitable water vapor measurements from the SMT were not available and some seeing measurements were not recorded in the data headers towards the end of the night.} 
\label{tab:December_Run_2}

\begin{center}       
\begin{tabular}{l|c|c|c|c|c|c|c} 
\textbf{Target} & SpT & Nods       & Exposures & Integration Time & Total Exposure & AM & seeing \\
\textbf{Name}   &      &            & per Nod   & per Exposure (s) & Time (s)       &     & (arcsec)         \\ \hline
HD 81212   &  F5 primary     & 2         & 60       & 0.98825          & 118.59  &  1.57 & 2.0 - 2.4  \\
HD 81212   &  F5 primary     & 2         & 60       & 0.98825          & 118.59  & 1.60 & 1.50  \\
HD 81212   &  F5 primary     & 2         & 60       & 0.98825          & 118.59  & 1.62 & 1.5 - 2.1  \\
HD 81212   &  F5 primary     & 2         & 60       & 0.98825          & 118.59  & 1.65 & 1.8 - 2.4  \\
HD 81212   &  F5 primary     & 2         & 60       & 0.98825          & 118.59  & 1.67 & 1.5 - 3  \\
HD 81212   &  F5 primary     & 2         & 60       & 0.98825          & 118.59  & 1.69 & 3 - 1.5  \\
HD 81212   &  F5 primary     & 2         & 60       & 0.98825          & 118.59  &  1.72 & 1.5 - 3  \\
\hline
Alpha Leo   &  B8     & 2         & 60       & 0.98825          & 118.59  & 1.39 & - \\
Alpha Leo   &  B8     & 2         & 60       & 0.98825          & 118.59  & 1.40 & -  \\

\end{tabular}
\end{center}
\end{table}

\begin{table}[ht]
\caption{Summary of targets observed for the first on sky run on 2023-Dec-27. HD 10453 is a binary, but the spectral type of the primary is listed. Precipitable water vapor measurements from the SMT were not available and some seeing measurements were not recorded in the data headers towards the end of the night.} 
\label{tab:December_Run_3}

\begin{center}       
\begin{tabular}{l|c|c|c|c|c|c|c} 
\textbf{Target} & SpT & Nods       & Exposures & Integration Time & Total Exposure & AM & seeing \\
\textbf{Name}   &      &            & per Nod   & per Exposure (s) & Time (s)       &     & (arcsec)         \\ \hline
HD 10453  &  F5 primary     & 2         & 60       & 0.98825          & 118.59  &  1.40 & 1.7-2.3\\
HD 10453  &  F5 primary     & 2         & 60       & 0.98825          & 118.59  &  1.40 & 2.5 - 3.5\\
HD 10453  &  F5 primary     & 2         & 60       & 0.98825          & 118.59  &  1.40 & 1.6 - 3.0\\
HD 10453  &  F5 primary     & 2         & 60       & 0.98825          & 118.59  &  1.39 & 1.59\\
HD 10453  &  F5 primary     & 2         & 60       & 0.98825          & 118.59  &  1.39 & 1.67\\
HD 10453  &  F5 primary     & 2         & 60       & 0.98825          & 118.59  &  1.39 & 1.84\\
HD 10453  &  F5 primary     & 2         & 60       & 0.98825          & 118.59  &  1.39 & 1.6-3.5\\
HD 10453  &  F5 primary     & 2         & 60       & 0.98825          & 118.59  &  1.39 & 2.0-3.0\\

\hline
Alpha Piscium   &  B8     & 2         & 60       & 0.98825          & 118.59  & 1.39 & - \\
Alpha Piscium  &  B8     & 2         & 60       & 0.98825          & 118.59  & 1.40 & -  \\

\end{tabular}
\end{center}
\end{table}

\begin{table}[ht]
\caption{Summary of standard stars observed for the first on sky run on 2023-Dec-29. Precipitable water vapor measurements from the SMT were not available during these observations.} 
\label{tab:December_Run_4}

\begin{center}       
\begin{tabular}{l|c|c|c|c|c|c|c} 
\textbf{Target} & SpT & Nods       & Exposures & Integration Time & Total Exposure & AM & seeing \\
\textbf{Name}   &      &            & per Nod   & per Exposure (s) & Time (s)       &     & (arcsec)         \\ \hline
Alpha Coronae Borealis &  A1     & 2         & 60       & 0.98825          & 118.59  &  1.44 &  1.0\\
Alpha Coronae Borealis &  A1     & 2         & 60       & 0.98825          & 118.59  &  1.42 &  1.0\\
Alpha Coronae Borealis &  A1     & 2         & 60       & 0.98825          & 118.59  & 1.41 & 1.1 \\
Alpha Coronae Borealis &  A1     & 2         & 60       & 0.98825          & 118.59  &  1.40 & 1.1 \\
Alpha Coronae Borealis &  A1     & 2         & 60       & 0.98825          & 118.59  & 1.39 & 1.0 \\
Alpha Coronae Borealis &  A1     & 2         & 60       & 0.98825          & 118.59  & 1.37 & 1.1 \\
Alpha Coronae Borealis & A1     & 2         & 60       & 0.98825          & 118.59  &  1.36 & 1.1 \\

\hline

Beta Ursae Majoris & A1     & 2         & 60       & 0.98825          & 118.59  & 1.13   & 1.3 \\
Beta Ursae Majoris & A1     & 2         & 60       & 0.98825          &118.59  & 1.13  & 1.2  \\
Beta Ursae Majoris & A1     & 2         & 60       & 0.98825          & 118.59  & 1.14  & 1.3  \\
Beta Ursae Majoris & A1     & 2         & 60       & 0.98825          & 118.59  & 1.14  & 1.4 \\
Beta Ursae Majoris & A1     & 2         & 60       & 0.98825          & 118.59  & 1.14  & 1.3 \\
Beta Ursae Majoris & A1     & 2         & 60       & 0.98825          & 118.59  & 1.14  & 1.3 \\
Beta Ursae Majoris & A1     & 2         & 60       & 0.98825          & 118.59  & 1.15  & 1.2 \\
Beta Ursae Majoris & A1     & 2         & 60       & 0.98825          & 118.59  & 1.15  & 1.2  \\

\hline

Alpha Leo & B8     & 2         & 60       & 0.98825          & 118.59  & 1.40   & 1.2 \\
Alpha Leo & B8     & 2         & 60       & 0.98825          & 118.59  & 1.41  & 1.3  \\
Alpha Leo & B8     & 2         & 60       & 0.98825          & 118.59  & 1.42   & 1.1 \\
Alpha Leo & B8     & 2         & 60       & 0.98825          & 118.59  & 1.43 & 1.0  \\
Alpha Leo & B8     & 2         & 60       & 0.98825          & 118.59  & 1.45  & 1.1 \\
Alpha Leo & B8     & 2         & 60       & 0.98825          & 118.59  & 1.46  &  1.0 \\
\end{tabular}
\end{center}
\end{table}

\section{Data Reduction} \label{sec:data reduction}
The spectral traces produced by NOMIC's grism are straight but slanted in the detector images by a few degrees. Following the procedures used in the reduction code REDSPEC for Keck/NIRSPEC (\citenum{2015ascl.soft07017K}), detector images are interpolated in two directions so spectral traces become parallel with the y-direction. Interpolation is not always necessary for very bright targets, but it is helpful when dimmer science targets need to be collapsed along the wavelength direction to confirm detection and for centering the target for spectral extraction.

The first step to getting fully rectified images is using the A-B nod pair-subtracted images to fit centroids along the two spectral traces in the image. Best fit lines to these traces are used to estimate the shift at each row along the x-direction to align the traces along the y-direction. The first rectification spatial map is created and applied to the A nod image. The rectified A nod image is then used to fit centroids along three, broad sky background lines to estimate the shift at each column to align the two traces in the wavelength direction for the second spectral rectification map. Both spatial and spectral rectification maps are applied to the average A and B nods separately, then the fully rectified A-B image is calculated. For dimmer targets, the rectification maps derived from a brighter target are used to interpolate the images. Excess sky background is removed by calculating the mean along each row and subtracting it from the fully rectified A-B image. The portions of the spectra impacted by the boundaries of NOMIC's detector channels are not interpolated over or removed prior to rectification, but masked out after extraction.

\subsection{Wavelength Solution}
No filter calibrations were taken on 2023-Sep-24, therefore those observations were wavelength calibrated by comparing the data's sky background to a smoothed reference sky background and the edges of the W10288-8E filter (7.5 $\mu$m and 13.5 $\mu$m). The reference sky background feature used was an emission peak at 9.57 $\mu$m in Earth's atmosphere that appeared in all data. A linear wavelength solution was fit through the three benchmarks for a wavelength solution. The wavelength solution of the science images is much finer than the true resolution of the grism. The extracted spectra are binned by at least 5 pixels to match the estimated resolution of about 200. Observations taken in December had a set of filter images taken with the grism (as shown in Figure~\ref{fig:footprint}, Right 6 panels), however the detector channel-dependent noise and biases makes estimating the full width at half maximums for each filter challenging. During the 2023-Dec-29 run, filter images were taken, but no dark frames were taken to mitigate detector channel artifacts. In the future, both dark frames and filter sets need to be taken during NOMIC observations to refine wavelength solution strategies.

\subsection{Spectral Extraction and Error Estimation}
The fully rectified mean A-B spectral images are collapsed by taking a weighted average along the spectral direction in order to find the centers of the spectral traces. A Gaussian is fit to the positive and negative traces profile to estimate their amplitude and width. A very wide 3$\sigma$ boxcar was used on all targets. For calibrator stars, the average extraction was 42 pixels wide. For dimmer targets, the average extraction width ranged from 35-38 pixels depending on seeing. The error was estimated by taking the standard deviation as a function of wavelength using a 50 pixel wide area away from the spectral traces. The error associated with each extracted trace is the square root of the extraction width multiplied by the standard deviation of the background as a function of wavelength.

\subsection{Telluric Correction and Relative Flux Calibration}
Calibrator stars are necessary to remove the response of the observatory, instrument, and Earth's atmosphere. All calibrator targets in this work are late B or early A type stars that have spectral responses similar to a blackbody across the mid-infrared to first order. Extracted target spectra are divided by extracted calibrator spectra and then multiplied by a blackbody function matching the temperature of the calibrator star.  The flux calibrations are left as relative and normalized because no photometry was taken to produce a reliable absolute flux calibration. In infrared astronomy beyond 3 $\mu$m, it is common for calibrations to be taken at similar airmasses and every one to two hours over the course of science observations. At longer wavelengths (10 $\mu$m), meeting these requirements for very bright targets to test a new instrument mode during low priority queue is difficult. There are mismatches in airmass among target-calibrator pairs that result in a poor telluric calibration in the spectra taken between 9.3-9.9 $\mu$m.

\section{Initial Mode Performance} \label{sec:mode performance}

\subsection{Wavelength Coverage and Resolution}
On-sky observations show that the wavelength coverage of NOMIC's low resolution mode is 8-13 $\mu$m. The approximate wavelength range of the W10288-8E filter covers 8-14 $\mu$m, but the usable wavelength coverage on-sky within Earth's atmospheric window stops at 13 $\mu$m. The signal of our brightest calibrator star Sirius plateaus at 12.8 $\mu$m and the science target did not get enough exposure time to get appreciable signal to noise ($>$ 5) beyond 12.5 $\mu$m. Given that the observations were taken in poor conditions, it is likely airmass and water vapor limits the atmospheric transmission from 12.5-13.0 $\mu$m. Atmospheric transmission simulations from European Southern Observatory's tool SKYCALC \footnote{\url{https://www.eso.org/observing/etc/bin/gen/form?INS.MODE=swspectr+INS.NAME=SKYCALC}} [\citenum{2012A&A...543A..92N}] shows transmission from 12.8-13.0 $\mu$m is low, but still improves with lower airmass and lower precipitable water vapor. The resolution of the grism was inferred by binning a model sky background to match the width between the peaks of the 9.57 $\mu$m emission feature. The best estimate at this time is a resolution of $\sim$200, which will be improved with observations in better conditions and science targets that are not featureless stellar spectra.

 \subsection{Sensitivity}

Lambda Persei is a 1.07 Jy source in the N-band based on the MATISSE catalog of sources (\citenum{2019MNRAS.490.3158C}). As a A1 spectral type star, its response should be close to a blackbody response. Each extracted A-nod spectrum of Lambda Persei is divided by the extracted A-nod of Sirius. The same is done with the extracted B-nods of Lambda Persei and Sirius. The telluric calibrated spectra are normalized by their median and then combined into one spectrum by a weighted average. The spectrum is binned by 5 pixels again for two reasons: 1) to match the estimated resolution of the grism and 2) smooth over ratio errors caused by differences in wavelength solutions of Lambda Persei and Sirius.

\begin{figure} [ht]
   \begin{center}
   \begin{tabular}{c} 
   \includegraphics[width=6.5in]{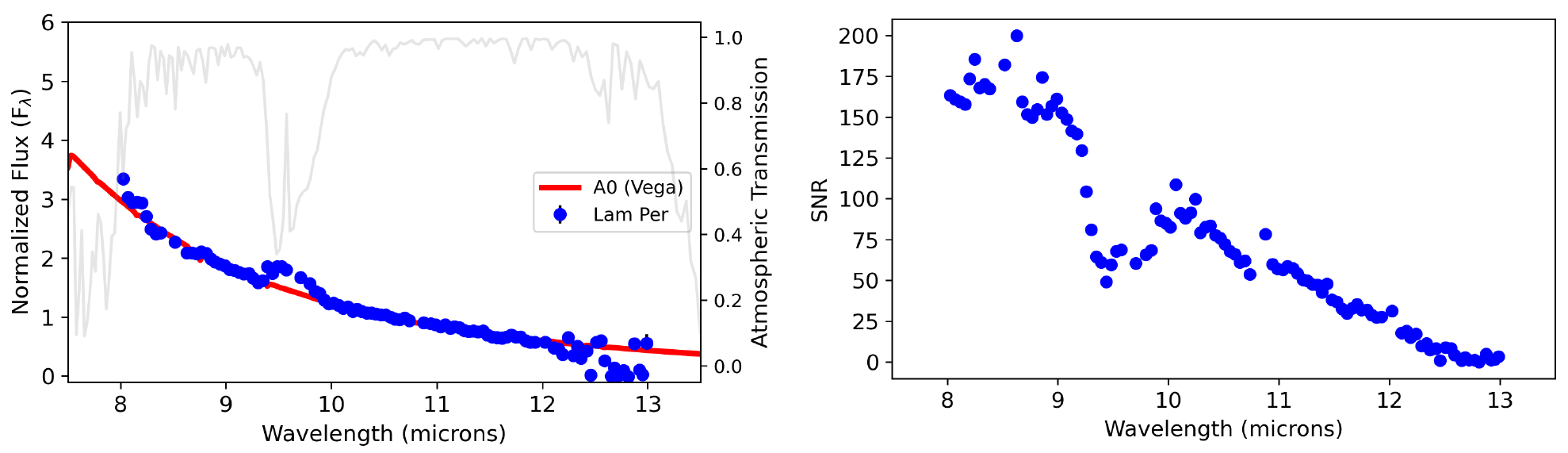}
   \end{tabular}
   \end{center}
   \caption[example]{ \label{fig:spectrum_example} LBTI/NOMIC spectrum of Lambda Persei (blue dots with black error bars) combining 15.8 minutes of total exposure time and calibrated with 2 minutes of observations of Sirius. The red line is a normalized spectrum of the idealized version of Vega (\citenum{2008AJ....135.2245R}, A0), which is the same spectral type as Lambda Persei. The atmospheric transmission is plotted in grey above the spectrum. The airmass difference between Lambda Persei and Sirius produces telluric errors in the final spectrum especially in regions of low transmission. The right panel shows the signal-to-noise of the weighted average for all extracted and telluric calibrated spectra of Lambda Persei. Outside of low transmission regions, NOMIC follows the expected A0 star spectrum even in poor conditions.}
\end{figure}

 NOMIC's spectroscopic mode was able to re-produce the spectrum of a standard A star in poor observational conditions. The reduced spectra of the Lambda Persei observations in Table~\ref{tab:September_Run} are shown in Figure~\ref{fig:spectrum_example}. Over about 16 minutes and including telluric overheads, a signal to noise of 150-180 was measured from 8-9 $\mu$m. From 10-12.5 $\mu$m the signal to noise fell linearly from 100 to 10. Within the low transmission region between 9-10 $\mu$m the signal to noise was 70. The signal to noise of Sirius was on average 1000 per pixel contributing very little to the error budget. The observed airmass discrepancy between Sirius and Lambda Persei created a bump due to poor telluric calibration in the reduced spectrum from 9.3 - 9.9 $\mu$m. Sirius was observed at a higher airmass, reducing the amount of counts measured within this wavelength region. This effect will be described further in the next section. The spectrum of Lambda Persei was compared against an idealized reference spectrum of Vega (A0) (\citenum{2008AJ....135.2245R}), showing that Lambda Persei's spectrum follows a blackbody.

\subsection{Effect of Airmass on Telluric Correction}
\begin{figure} [ht]
   \begin{center}
   \begin{tabular}{c} 
   \includegraphics[width=6.5in]{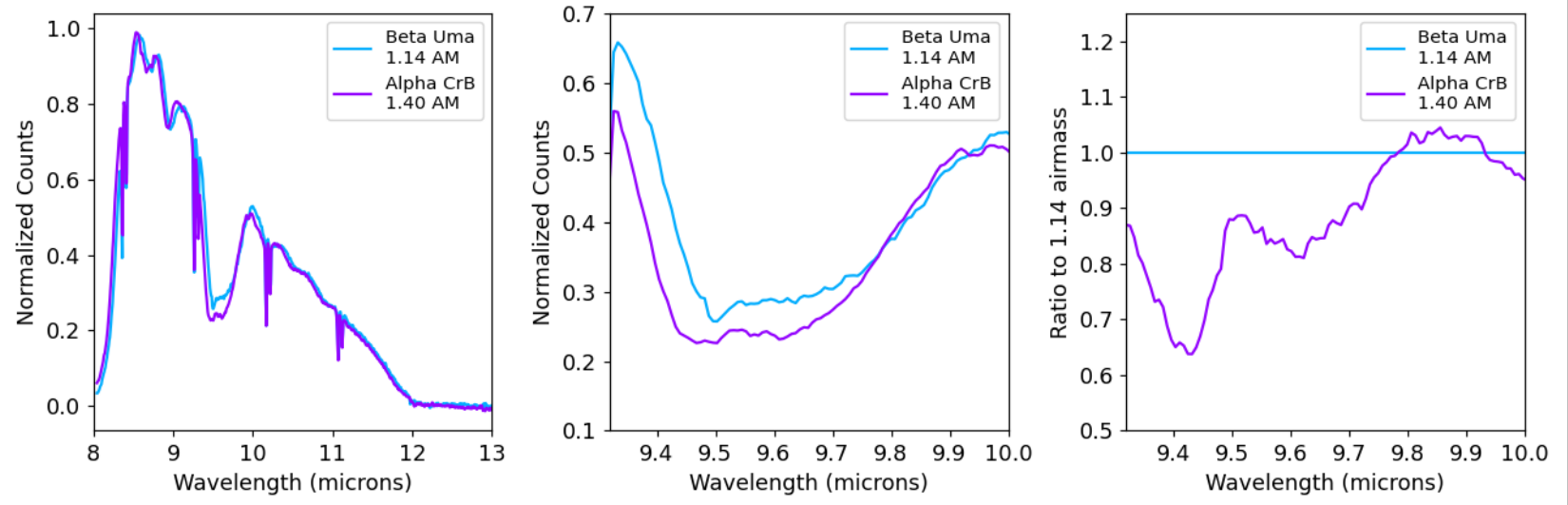}
   \end{tabular}
   \end{center}
   \caption[example]{ \label{fig:airmass_test} Left: Normalized average raw counts of Alpha Coronae Borealis (purple) and Beta Ursae Majoris (blue), two stars with the same spectral type (A1) taken at different airmasses. The dips near 8.3, 9.29, 10.2, 11.1 $\mu$m are caused by the boundaries of NOMIC's detector channels and are not astronomical. Middle: Zoom in of the left figure between 9.3-10 $\mu$m, where airmass impacts the count response the most. Right: Ratio of the counts compared to the 1.14 airmass average Beta Ursae Majoris.}
\end{figure}

Increased airmass reduces the measured counts of targets between 9.3-9.8 $\mu$m. Observations from Table ~\ref{tab:December_Run_4} were reduced and extracted but not telluric calibrated. For Alpha Coronae Borealis and Beta Ursae Majoris the average extracted nod had a signal to noise per pixel of 50-100. Alpha Leo's average signal to noise for an extracted nod was 75-200 per pixel. Only Alpha Coronae Borealis and Ursae Majoris were used in comparisons because they are the same spectral type (A1). Each star has the average calculated between all normalized nods and they are plotted in Figure~\ref{fig:airmass_test}. Based on our very bright targets, a .26 increase in airmass from 1.14 to 1.40 correlates to the reduction of calibrator response by about 10$\%$ at 9.5 $\mu$m. No bright calibrator stars were taken to provide a correction for the Lambda Persei data. 

\subsection{Binary Observations}
\begin{figure} [ht]
   \begin{center}
   \begin{tabular}{c} 
   \includegraphics[width=6.5in]{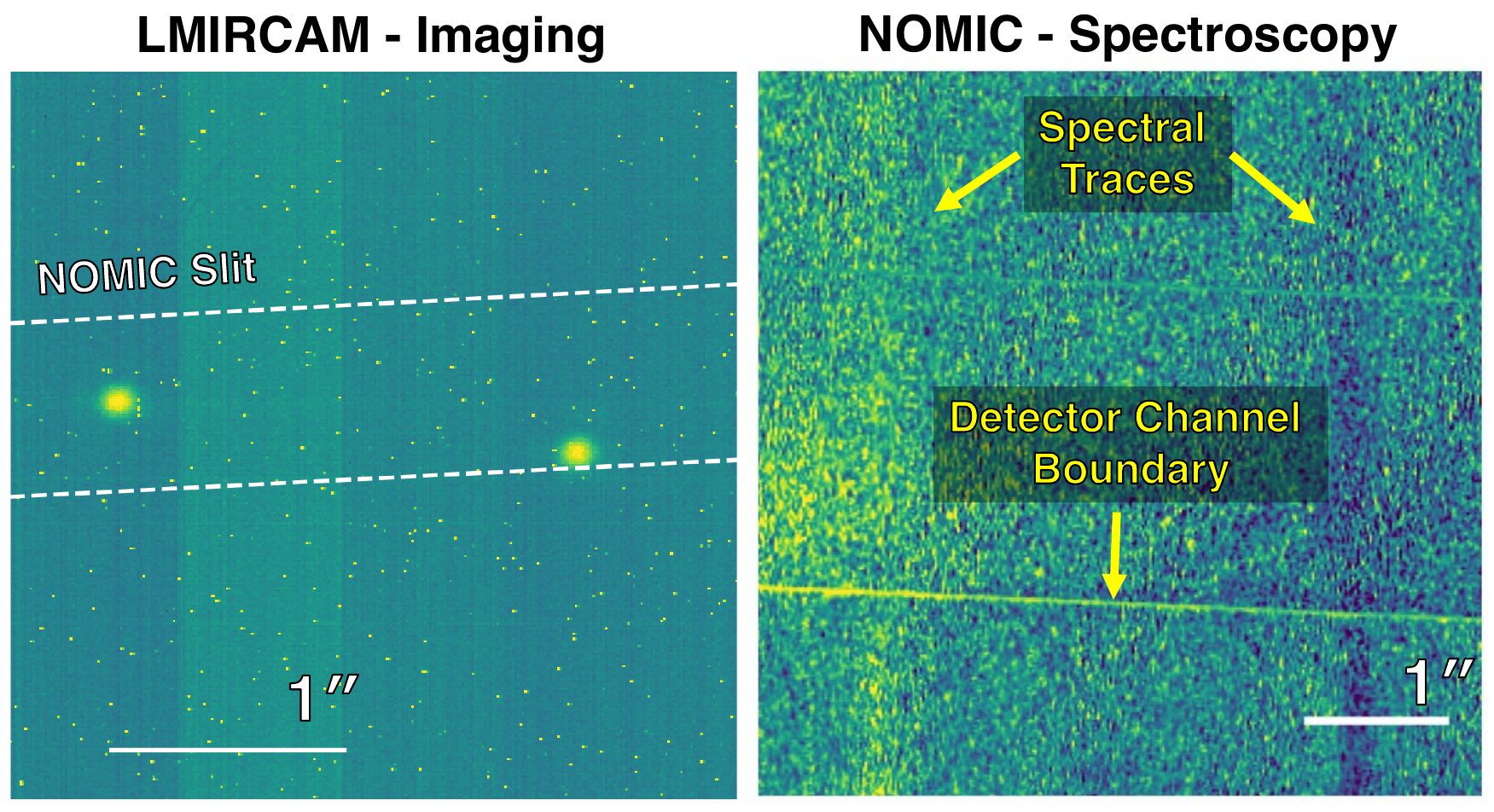}
   \end{tabular}
   \end{center}
   \caption[example]{ \label{fig:HD81212} Left: Raw detector image of HD 81212 A and B from LMIRCam. The stamp image is 350 by 350 pixels wide. The size (not the exact location) of the NOMIC slit is overplotted on the LMIRCam image with dashed, white lines. Right: A-B nod subtracted spectral image of the system. A 290 by 290 pixel subset is plotted. Only one positive and negative trace from a single star is easily visible in the NOMIC spectral image. A white line marking the length of 1$^{\prime\prime}$ is displayed on both instrument images.}
\end{figure}

\begin{figure} [ht]
   \begin{center}
   \begin{tabular}{c} 
   \includegraphics[width=6.5in]{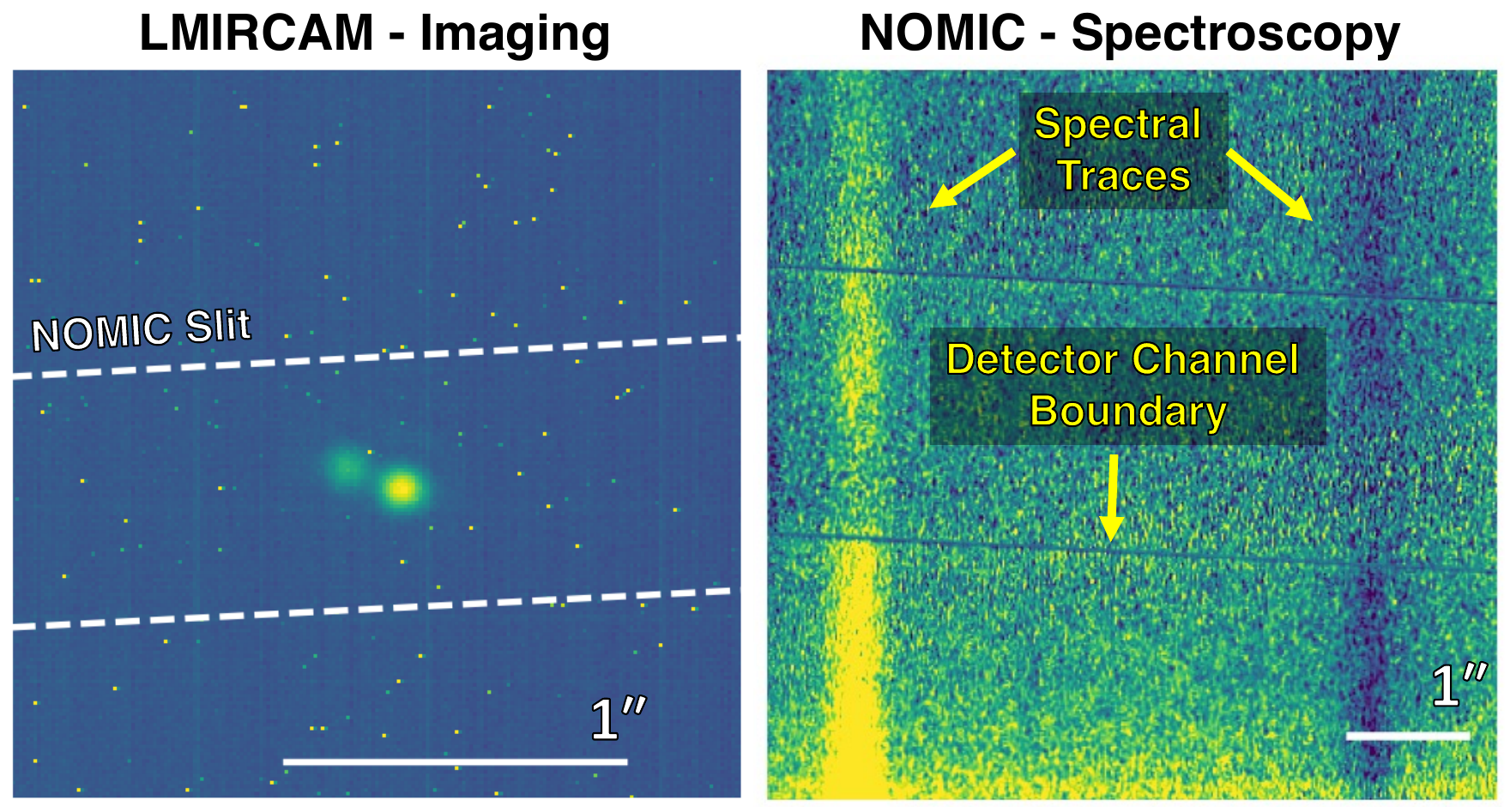}
   \end{tabular}
   \end{center}
   \caption[example]{ \label{fig:HD10453} Left: Raw detector image of HD 10453 A and B from LMIRCam. The stamp image is 200 by 200 pixels wide. The size (not the exact location) of the NOMIC slit is overplotted on the LMIRCam image with dashed, white lines. Right: A-B nod subtracted spectral image of the system. A 350 by 350 pixel subset is plotted. Only one positive and negative trace from a single star is easily visible in the NOMIC spectral image. A white line marking the length of 1$^{\prime\prime}$ is displayed on both instrument images.}
\end{figure}

Two binary systems were observed to understand how well NOMIC can spatially resolve targets. The observations taken on 2023-Dec-26 and 2023-Dec-27 of HD 81212 and HD 10453 were reduced from raw data but not extracted and telluric calibrated. Only a few images of the HD 81212 system were taken with LMIRCam at L-band to center the target in NOMIC's slit. The stars in the HD 81212 system were separated far enough on sky during observations that both stars should be spatially resolved by NOMIC. Only one negative and one positive stellar trace pair can be easily seen in the reduced NOMIC spectral images. Poor positioning of the NOMIC slit could have caused one of the HD 81212 components to fall out of the slit's field of view. The LMIRCam images and NOMIC spectroscopy of the HD 81212 system are shown in Figure~\ref{fig:HD81212}. Observations of the HD 10453 system consisted of contemporaneous LMIRCam L-band imaging and NOMIC spectroscopy over all observation blocks listed in Table~\ref{tab:December_Run_3}. The two stars in the system were observed much closer than expected with a x-direction separation of $\sim$.16$^{\prime\prime}$, less than the $\sim$.25$^{\prime\prime}$ diffraction limit at 10 $\mu$m. The two stars are both in the slit and blended in the reduced spectral images in Figure~\ref{fig:HD10453}.

\section{Future Work and Discussion} \label{sec:discussion}

\subsection{Observational Strategy and Plans}
Our technical time observations using NOMIC's spectroscopic mode can reproduce the spectrum of a bright A-star which resembles a blackbody despite poor conditions. There were issues with the telluric calibration within low atmospheric transmission areas, but in the future this can be mitigated or at least characterized by choosing better calibration targets. Based on initial on-sky testing of the mode, the minimum set of calibrations suggested for NOMIC spectroscopy are 1) two nod pairs of a bright calibrator star 2) dark images at the integration length of science observations and 3) pictures of the grism with filters as shown in Figure~\ref{fig:footprint}. The frequency at which calibrator stars need to be observed will be explored when science observations longer than several hours are done. 

The next step for assessing this instrument mode on sky is observing targets that have a scientific and technical purpose during Fall 2024. The first target will be a Solar System gas giant. All Solar System gas giant planets have hydrocarbons appearing in emission across the mid-infrared due to methane photochemistry (\citenum{2023RemS...15.1811R}). The ice giants Neptune and Uranus lack dedicated orbiter missions and are dependent on telescopes for seasonal observations of these systems. Space-based observatories often have the sensitivity to measure these spectral features, but lack the spatial resolution to associate spectral features with distinct surface features like bands or spots. Ground-based observatories often have the advantage of easier access to execute time-domain studies of spectral features over seasons. 

The next two targets will be a debris disk and a star with a substellar/planet companion. A bright debris disk like Vega or another HOSTS survey target (\citenum{2020AJ....159..177E}) will be observed to understand if and how well spatial changes in grain chemistry can be measured in a debris disk. Lastly, another star with a close, substellar companion will be chosen to refine combined LMIRCam imaging and NOMIC spectroscopic observational strategies and reduction techniques. Understanding the onset temperature and evolution of silicate clouds along with the detection of ammonia (see Figure~\ref{fig:HR_diagram_and_spectra}) is crucial for the characterization of the atmospheres of directly-imaged exoplanets. In this proceeding, the background mean at each wavelength was used for excess noise subtraction, but a principal component analysis (PCA)-based technique used previously on NOMIC imaging data will be tested in the future to improve signal-to-noise on science targets (\citenum{2024arXiv240518043R}).

\subsection{Guided Spectral Stacking}
 For the data reduction presented, sources were found in nod-subtracted NOMIC spectral images by collapsing along the wavelength direction. Centering extraction apertures on dim companions could be improved and optimized by measuring relative offsets in images taken at another wavelength. LBTI has the unique capability to operate LMIRCam between 1-5 $\mu$m simultaneously with NOMIC (8-13 $\mu$m). This capability is powerful for efficiently following up known sources with longer wavelength spectroscopy. Using LMIRCam images to guide spectral stacking will be revisited for Fall 2024 observations. Long wavelength observations to find and spectroscopically characterize directly imaged planets are time intensive on both 10- and 30-meter class observatories. Developing future mid-infrared instruments that operate simultaneously with other modes is important for maximizing science and consistently maintaining 10 $\mu$m capabilities at observatories.

\subsection{Impact of Current Mid-IR Detectors and Planned Upgrades}
In addition to sky background, the choice of long-wavelength detector significantly influences the sensitivity and performance of mid-infrared instruments. We discuss how existing mid-infrared detectors could impact mid-infrared spectroscopy on the LBT and potential higher-resolution modes that can be developed and scaled up to 30-meter class observatories. There are plans to upgrade the current Si:As Impurity Band Conduction Aquarius array developed by Raytheon Vision Systems on NOMIC (\citenum{2012SPIE.8453E..12I, 2014SPIE.9154E..1JI}) to the HgCdTe GeoSnap array developed by Teledyne Imaging Sensors (TIS) (\citenum{2020gbti.confE...4A, 2023AN....34430103L, 2024arXiv240520440B}), which has been used on-sky for MMT/MIRAC-5 (\citenum{2022SPIE12184E..1UB}) and is planned for use on ELT/METIS (\citenum{2020SPIE11454E..1YG}). TIS has extended the wavelength cutoff out to 15 $\mu$m for HgCdTe HAWAII-xRG arrays (\citenum{2019JATIS...5c6005C, 2020JATIS...6a1004C}), which have a similar readout architecture to HxRG arrays commonly used at shorter wavelengths. The development of long-wavelength HxRG arrays was primarily driven by the requirements for the space-based NEO Surveyor mission (\citenum{2022SPIE12191E..2AR}). The Aquarius arrays have relatively low dark current, but exhibit Excess Low Frequency Noise (ELFN, \citenum{1998SPIE.3379..361A}) requiring chopping to mitigate. Long wavelength HAWAII arrays have low dark current, but do not have sufficient well depth to accommodate the sky background without sacrificing field of view (\citenum{10.1117/12.2594915}). The GeoSnap has better overall noise properties compared to the Aquarius array, enabling better sensitivity on NOMIC to directly image nearby habitable zone planets (\citenum{2021SPIE11823E..0GW, 2022SPIE12183E..02E}). The GeoSnap is ideal for imaging, but the comparatively high dark current becomes an issue for spectroscopy.

\begin{figure} [ht]
   \begin{center}
   \begin{tabular}{c} 
   \includegraphics[width=6.5in]{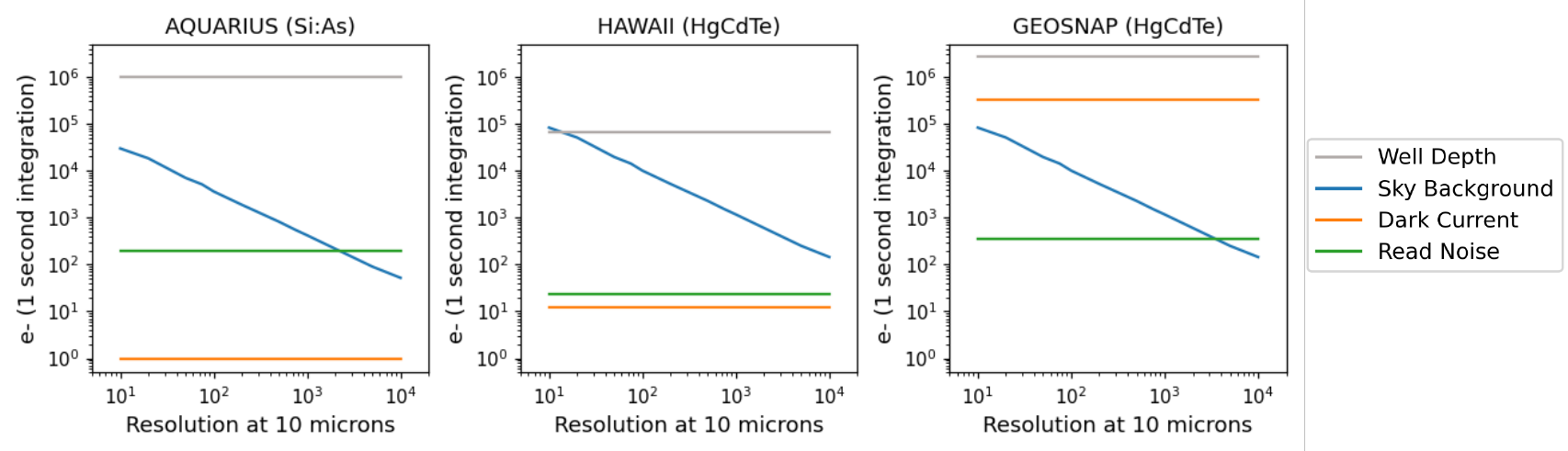}
   \end{tabular}
   \end{center}
   \caption[example]
   { \label{fig:detector_parameters_8.4m} The response of sky background (blue), dark current (orange), and read noise (green) plotted against spectral resolution for three different mid-infrared detectors: Aquarius, HAWAII, and GeoSnap arrays. The set up assumes NOMIC's pixel scale on a 8.4 meter aperture. The well depth of each detector is plotted as a grey line. If any noise source lies above that line, the detector is saturated. The detector's behavior is ideal when the noise is sky background limited.}
\end{figure}

\begin{figure} [ht]
   \begin{center}
   \begin{tabular}{c} 
   \includegraphics[width=6.5in]{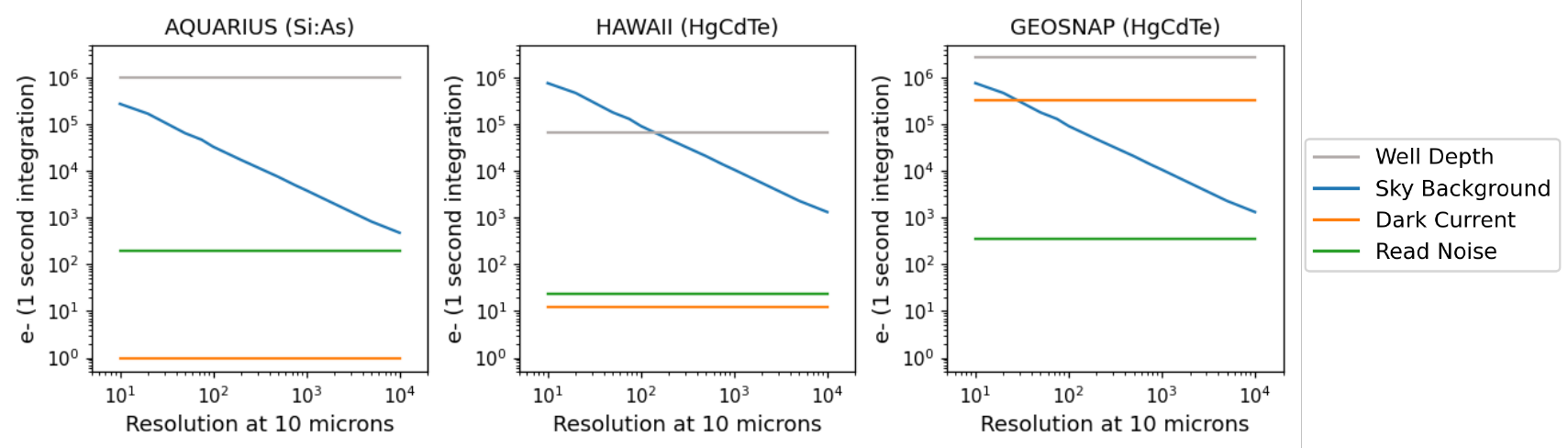}
   \end{tabular}
   \end{center}
   \caption[example]{ \label{fig:detector_parameters_25.4m} The plots uses the same labeling as Figure~\ref{fig:detector_parameters_8.4m}, but this time NOMIC is placed behind a 25.4 meter aperture.}
\end{figure}

Detector development is required to achieve a more sensitive long-wavelength array suitable for both imaging and spectroscopy. Using the published parameters of the Aquarius, HAWAII, and GeoSnap arrays we estimate the contribution of atmospheric background versus spectral resolution at 10 $\mu$m and compare it to dark current and read noise. The detector properties are summarized in Table~\ref{tab:detector_parameters}. A flat detector quantum efficiency of 70$\%$ is assumed with a filter transmission of 80$\%$. The sky transmission and background emission models adopted are based on the Maunakea site assuming an observation airmass of 1.5 and water vapor column of 1.6mm \footnote{\url{https://www.gemini.edu/observing/telescopes-and-sites/sites}}. The NOMIC plate scale of 17.9 mas per pixel is also assumed. Two different telescope apertures, 8.4 meters (one side of the LBT) and 25.4 meters (Giant Magellan Telescope) were explored. The estimated sky background after a 1-second integration at different resolutions is plotted for each detector in Figures~\ref{fig:detector_parameters_8.4m} and~\ref{fig:detector_parameters_25.4m}. For the current version of NOMIC on the LBT, the Aquarius array produces less dark current than the sky background per pixel at resolutions between R $\sim$10 - 10,000. At 1-second integrations, the Aquarius array would be read noise limited at a resolution of $\sim$2000 or higher. For the HAWAII array, the dark current is never larger than the sky background at any resolution between R $\sim$10 - 10,000 however, the sky background emission would fill the entire well depth capacity near a resolution of 10 or less. The dark current of the GeoSnap is larger than the sky background at resolutions of $\sim$10 and higher. The upgrade of the Aquarius array to the GeoSnap will improve broadband imaging on NOMIC, but will also contribute excess dark current noise that is comparable to the sky background in NOMIC's R$\sim$200 spectroscopy mode. At standard operating temperature, the dark current value of the GeoSnap used in our calculations is relatively high, but could be improved by operating the detector at colder temperatures.

If NOMIC were placed behind a larger aperture telescope with the same pixel scale, the amount of sky photons measured per pixel increases while the dark current per pixel remains the same. This increases the spectral resolution where observations are primarily sky background limited. Considering a 25.4 meter aperture, the GeoSnap would be sky background limited up to a resolution of $\sim$30. The dark current does not exceed the sky background for both Aquarius and HAWAII arrays from R $\sim$10 - 10,000, but they would still have ELFN noise and limited well capacity respectively. These results are shown in Figure~\ref{fig:detector_parameters_25.4m}. To enable more sensitive medium to high resolution, mid-infrared spectroscopic capabilities on 30-meter class facilities, higher well depth, lower dark current arrays need to be developed and tested.

\begin{table}[ht]
\caption{Published mid-infrared detector parameters used for noise estimations. Aquarius detector properties are primarily adopted from Table 1 in Reference \citenum{2014SPIE.9154E..1JI}. The well depth of the Aquarius array is adopted from high gain NOMIC measurements (\citenum{2014SPIE.9147E..1OH}). The mid-infrared HAWAII array parameters are adopted from measurements of wafer $\#$18367 (250mV bias, 28K) in Reference~\citenum{2019JATIS...5c6005C}, the read noise of the HAWAII arrays is determined from other wafers in the same study.  The GeoSnap parameters are adopted from Reference \citenum{2024arXiv240520440B}, see also Table 1 of Reference \citenum{2023AN....34430103L}.} 
\label{tab:detector_parameters}

\begin{center}       
\begin{tabular}{l|c|c|c} 
\textbf{Detector Property} &   Aquarius (Si:As)  & HAWAII (HgCdTe) & GeoSnap (HgCdTe) \\ \hline
Operating Temperature (K)    &   7-8    & 28     & 45  \\
Pixel Size ($\mu$m)    &   30    & 18     & 18  \\
Well Depth  (e-)       &   1.0 $*$ 10$^6$ & 6.7 $*$ 10$^4$ & 2.75 $*$ 10$^6$ \\
Dark Current (e-/pix/s)&  1             & 12           & 330,000 \\
Read Noise (e-)        &   200          & 23           & 360 \\

\end{tabular}
\end{center}
\end{table}

\section{Summary}
The Large Binocular Telescope Inteferometer's  long-wavelength camera NOMIC is capable of low resolution (R $\sim$200) spectroscopy over 8-13 $\mu$m and was able to measure the response of a standard blackbody source. Lambda Persei, which is a $\sim$ 1 Jansky N-band source, reached a signal to noise of 100 at 10 $\mu$m in 17.8 minutes, including calibration overheads. The wavelengths between 9.3-9.9 $\mu$m are very sensitive to telluric differences between the science target and calibrator. A 0.3 increase in airmass, led to a 10 percent reduction in counts over that region. This is important to consider when searching for molecules that have significant absorption across these wavelengths (example: ozone or silicate clouds). Ammonia, which is prominent in cooler gas giants falls outside of this region (see Figure~\ref{fig:HR_diagram_and_spectra}). Observations conducted during higher priority queue in better conditions will hopefully improve constraints on the mode's performance. The planned detector upgrades for NOMIC will likely impact the sensitivity, but lower operating temperatures can be explored to reduce the GeoSnap's dark current. Beyond the LBT, more detector development is necessary to enable higher sensitivity for sky background limited, medium to high resolution spectroscopy.

\acknowledgments 
The results reported herein benefited from collaborations and/or information exchange within NASA's Nexus for Exoplanet System Science (NExSS) research coordination network sponsored by NASA's Science Mission Directorate.

B.E.M. was supported by the Heising–Simons Foundation 51 Pegasi b Postdoctoral Fellowship. B.E.M would like to acknowledge the song “Rhythm Nation" by Janet Jackson for getting her through hard sets at the gym and being just as relevant today as it was when first released. 

\bibliography{report} 
\bibliographystyle{spiebib} 

\end{document}